\documentclass{article}
\usepackage{romp}

\usepackage{latexsym}
\usepackage{amssymb}
\usepackage{amsmath}
\def\cl{{\cal C}\!\ell}

\def\R{{\mathbb R}}
\def\C{{\mathbb C}}
\def\F{{\mathbb F}}

\def\diag{{\rm diag}}
\def\U{{\rm U}}
\def\Or{{\rm O}}
\def\SU{{\rm SU}}
\def\su{{\mathfrak su}}
\def\Mat{{\rm Mat}}
\def\tr{{\rm tr}}
\def\det{{\rm det}}
\def\Even{{\rm Even}}
\def\Odd{{\rm Odd}}
\def\T{{\rm T}}
\def\Tr{{\rm Tr}}
\def\Pin{{\rm Pin}}
\def\w{{\rm w}}
\def\proof{\medbreak\noindent{\bf Proof}}

\title{General solutions of one class of field equations}
\author{N.~G.~Marchuk \thanks{\, This work was supported by Russian Science Foundation (project RSF 14-50-00005, Steklov Mathematical Institute).} \\ Steklov Mathematical Institute,\\ Russian Academy of Sciences, Gubkina St. 8, 119991 Moscow, Russia \\ e-mail: nmarchuk@mi.ras.ru  \\[2ex]
         D.~S.~Shirokov
                      \\ National Research University Higher School of Economics,\\ Myasnitskaya str. 20, 101000, Moscow, Russia \\ e-mail: dshirokov@hse.ru\\Kharkevich Institute for Information Transmission Problems,\\ Russian Academy of Sciences, Bolshoy Karetny per. 19, 127051, Moscow, Russia\\e-mail: shirokov@iitp.ru }

\begin{document}

\maketitle
\begin{abstract}
    We find general solutions of some field equations (systems of equations) in pseudo-Euclidian spaces (so-called primitive field equations).  These equations are used in the study of the Dirac equation and Yang-Mills equations. These equations are invariant under orthogonal $\Or(p,q)$ coordinate transformations and invariant under gauge transformations, which depend on some Lie groups.  In this paper we use some new geometric objects - Clifford field vector and an algebra of h-forms which is a generalization of the algebra of differential forms and the Atiyah-K\"{a}hler algebra.
\end{abstract}


\section{Introduction}

In physics field equations describe physical fields and (using quantization) elementary particles. The following equations are fundamental relativistic  field equations: Maxwell's equations (1862), the Klein-Gordon-Fock equation (1926), the Dirac equation (1928), Yang-Mills equations (1954). These equations are considered in Minkowski space $\R^{1,3}$, they are invariant under Lorentz coordinate transformations. They are also invariant under certain unitary gauge transformations.

In this paper we consider a class of so-called  {\em primitive field equations} (systems of equations) (\ref{nik:eq}). These equations are considered  in pseudo-Euclidian spaces $\R^{p,q}$ and have different Lie groups of gauge symmetry. We find general solutions of primitive field equations corresponding to a wide class of gauge Lie groups. A partial case of such equations was considered in 1930-1939 in the theory of Dirac equation on curved pseudo-Riemannian manifolds of signature (1,3). Namely, it is a condition of generalized covariant constancy of $\gamma$-matrices, which is gauge invariant w.r.t. the spinor Lie group ${\rm Spin}(1,3)$ (see, for example, \cite{Mitsk} formula (4.5.13)).

In the theory of model field equations developed by authors in a series of papers and in monograph \cite{mybook:eng} there arise necessity to solve primitive field equations in pseudo-Euclidean space $\R^{1,3}$  with various gauge Lie groups (see Section 2). Also, primitive field equations are used in the theory of Yang-Mills equations. Namely, we present a new class of gauge-invariant solutions of  Yang-Mills equations, which correspond to solutions of  primitive field equations (see \cite{TrMIAN}).

In Section 2 of this paper we consider the Dirac equation and make some important notes about the Lie group $\SU(2,2)$ in connection with the Dirac equation. As a result of these notes we get new system of equations (\ref{gamC:eq1}), (\ref{gamC:eq2}). Equation (\ref{gamC:eq1}) is a special case of primitive field equations and we study these equations in the next sections of the paper in pseudo-Euclidian spaces $\R^{p,q}$.

In Section 3 we discuss some known facts about Clifford algebras. We actively use tensor fields with values in Clifford algebra.
Also we discuss some Lie algebras in Clifford algebra, especially Lie algebras $\w(\cl(p,q)))$. Results about Lie subalgebras of the Lie algebras $\w(\cl(p,q)))$ of pseudo-Unitary Lie group are our original results published in \cite{mybook:eng, MarShir:book}.

In Section 4 we present original results about projection operators and contractions in Clifford algebras. We use these results in Section 7.

In Section 5 of this paper we present some new geometric objects - Clifford field vector and an algebra of h-forms which is a generalization of the algebra of differential forms and the Atiyah-K\"{a}hler algebra \cite{atiyah, kahler}\footnote{We combine the technique of the Dirac gamma matrices and the technique of differential forms, in particular, the Atiyah-K\"{a}hler algebra of differential forms.}. These objects are helpful for consideration of some problems related to field theory equations.

In Sections 6 and 7 we consider a primitive field equation and present original results about its gauge symmetry and about general solutions of this equation.

Note that all considerations of this paper are valid for general pseudo-euclidian metric of signature $(p,q)$ and, in particular, for the Lorentzian metric $(+, -, -, -)$. Results of the paper can be understood either on the base of Dirac gamma matrices or on the base of Clifford algebras.

\section{A new view on the Dirac equation and $\gamma$-matrices}

Consider the Dirac equation for an electron in the Minkowski space $\R^{1,3}$ with coordinates $x^\mu$, $\mu=0,1,2,3$
($\partial_\mu=\partial/\partial x^\mu$ -- partial derivatives)
 \begin{equation}
i\gamma^\mu(\partial_\mu\psi-i a_\mu\psi)-m\psi=0,
\label{Dirac:eq:el}
\end{equation}
where $\gamma^\mu$ are 4 complex square matrices of order 4 satisfying conditions
\begin{eqnarray}
\partial_\mu\gamma^\nu &=& 0,\label{gam:eq1}\\
\gamma^\mu\gamma^\nu+\gamma^\nu\gamma^\mu &=&
2\eta^{\mu\nu}I,\label{gam:eq2}
\end{eqnarray}
where $\eta=\|\eta^{\mu\nu}\|=\diag(1,-1,-1,-1)$,
$I$ is the identity matrix of order 4, $a_\mu=a_\mu(x)$ is a covector potential of electromagnetic field, $\psi=\psi(x)$ is a Dirac spinor (column of four complex functions $\psi : \R^{1,3} \to \C^4$), $i$ is the imaginary unit, $m$ is a real number (mass of electron).

In the theory of the Dirac equation it is assumed that we have a fixed set of matrices $\gamma^\mu$ that satisfy conditions (\ref{gam:eq1}), (\ref{gam:eq2}) and the condition\footnote{This condition is required when we consider bilinear covariants of the Dirac spinors.} for Hermitian conjugated matrices
\begin{equation}
(\gamma^\mu)^\dagger=\gamma^0\gamma^\mu\gamma^0.\label{gam:herm}
\end{equation}

Matrices $\gamma^\mu$ satisfying conditions
(\ref{gam:eq1}), (\ref{gam:eq2}), (\ref{gam:herm}) are defined up to a similarity transformation with a unitary matrix
$U\in\U(4)$, i.e. matrices
\begin{equation}
\acute\gamma^\mu=U^{-1}\gamma^\mu U,\quad\hbox{where}\quad U^{-1}=U^\dagger\label{gam:U}
\end{equation}
satisfy the same conditions
(\ref{gam:eq1}), (\ref{gam:eq2}), (\ref{gam:herm}).

In particular, matrices $\gamma^0, \gamma^1, \gamma^2, \gamma^3$ in the Dirac representation satisfy these conditions and the matrix $\gamma^0$ is diagonal $\gamma^0=\diag(1,1,-1,-1)$. This matrix $\gamma^0$ changes under unitary transformation (\ref{gam:U}).

Denote $\beta=\diag(1,1,-1,-1)$ and  consider Lie group $\SU(2,2)$ of special pseudo-unitary matrices and its real Lie algebra $\su(2,2)$ (see \cite{Cornw})
\begin{eqnarray*}
\SU(2,2) &=& \{S\in\Mat(4,\C) : S^\dagger\beta S=\beta,\ \det\,S=1\},\\
\su(2,2) &=& \{s\in\Mat(4,\C) : \beta s^\dagger\beta=-s,\
\tr\,s=0\},
\end{eqnarray*}
where $\Mat(4,\C)$ is the algebra of complex matrices of order 4. Dirac gamma matrices $\gamma^\mu$ satisfy (\ref{gam:herm}) and $\tr\,\gamma^\mu=0$, therefore
\begin{equation}
i\gamma^\mu\in\su(2,2).\label{gam:su22}
\end{equation}

We may consider conditions (\ref{gam:eq1}), (\ref{gam:eq2}) together with condition (\ref{gam:su22}) and allow a similarity transformation
\begin{equation}
i\gamma^\mu\to i\acute\gamma^\mu=S^{-1}i\gamma^\mu S\label{gam:S}
\end{equation}
with matrix $S\in\SU(2,2)$, which preserves
(\ref{gam:eq1}), (\ref{gam:eq2}) and (\ref{gam:su22}).

If we consider conditions (\ref{gam:eq1}), (\ref{gam:eq2}) as equations for matrices $\gamma^\mu$ with condition
(\ref{gam:su22}), then we can consider transformation (\ref{gam:S}) as a global symmetry (it does not depend on $x\in\R^{1,3}$) of this system of equations.

Now we change equations (\ref{gam:eq1}) and obtain a system of equations with local (gauge) symmetry with respect to the pseudo-unitary group $\SU(2,2)$.

Namely, consider the following system of equations \cite{mybook:eng}:
\begin{eqnarray}
\partial_\mu\gamma^\nu - [C_\mu,\gamma^\nu] &=& 0,\label{gamC:eq1}\\
\gamma^\mu\gamma^\nu+\gamma^\nu\gamma^\mu &=&
2\eta^{\mu\nu}I,\label{gamC:eq2}
\end{eqnarray}
where $i\gamma^\mu=i\gamma^\mu(x)$ and $C_\mu=C_\mu(x)$ are smooth functions of $x\in\R^{1,3}$ with values in the Lie algebra $\su(2,2)$. The system of equations (\ref{gamC:eq1}), (\ref{gamC:eq2}) is invariant under the local (gauge) transformation
\begin{eqnarray}
i\gamma^\mu &\to& i\acute\gamma^\mu=S^{-1}i\gamma^\mu S,\label{gam:Sx}\\
C_\mu &\to& \acute C_\mu=S^{-1}C_\mu S-S^{-1}\partial_\mu
S,\label{C:Sx}
\end{eqnarray}
where the matrix $S=S(x)$ is a function of $x\in\R^{1,3}$ with values in the Lie group $\SU(2,2)$.

We consider system of equations (\ref{gamC:eq1}), (\ref{gamC:eq2}) as a new field equation (system of equations). We call this equation {\em a primitive field equation}. Let us analyze this equation in pseudo-Euclidian spaces. We use a formalism of Clifford algebras because, in our opinion, this formalism is the most convenient for this task.

\section{Clifford algebras}

Consider real $\cl^\R(p,q)$ or complexified $\cl(p,q)=\C\otimes\cl^\R(p,q)$ (see \cite{Lounesto}) Clifford algebra with $p+q=n$, $n\geq1$. Note that $\cl^\R(p,q)\subset \cl(p,q)$. When our argumentation is applicable to both cases, we write $\cl^\F(p,q)$, implying that $\F=\R$ or $\F=\C$. The construction of Clifford algebra is discussed in details in \cite{Lounesto, MarSh2008, MarShir:book}.

Let $e$ be the identity element and let $e^a$, $a=1,\ldots,n$ be generators\footnote{We use notation from \cite{Benn:Tucker} (see, also \cite{mybook:eng}). Note that there exists another notation instead of $e^a$ - with lower indices. But we use upper indices because we take into account relation with differential forms. Note that $e^a$ is not exponent.} of the Clifford algebra $\cl^\F(p,q)$,
\begin{equation}
e^a e^b+ e^b e^a=2\eta^{ab}e,\label{cond}
\end{equation}
where $\eta=||\eta^{ab}||=||\eta_{ab}||$ is the diagonal matrix with $p$ pieces of $+1$ and $q$ pieces of $-1$ on the diagonal. Elements
\begin{equation}
e^{a_1\ldots a_k}=e^{a_1}\cdots e^{a_k},\qquad a_1<\cdots<a_k,\,k=1,\ldots,n,\label{basis}
\end{equation}
together with the identity element $e$ form the basis of the Clifford
algebra. The number of basis elements is equal to $2^n$.

Any element $U$ of the Clifford algebra $\cl^\F(p,q)$ can be expanded in the basis:
\begin{eqnarray}
U=ue+u_a e^a+\sum_{a_1<a_2}u_{a_1 a_2}e^{a_1
a_2}+\cdots+u_{1\ldots n}e^{1\ldots n},\label{decompos}
\end{eqnarray}
where $u, u_a, u_{a_1 a_2},\ldots, u_{1\ldots n}$ are real or complex numbers (in the respective cases $\cl^\R(p,q)$ or $\cl(p,q)$).

Vector (real or complex) subspaces spanned on basis elements $e^{a_1\ldots a_k}$ labeled by ordered multi-indices of length $k$ are denoted by $\cl^\F_k(p,q)$, $k=0,\ldots, n$. Elements of the subspace $\cl^\F_k(p,q)$ are called {\em elements of rank}\footnote{There is a difference in notation in literature. We use term ``rank'' and notation $\cl_k(p,q)$ because we take into account a relation with differential forms, see \cite{mybook:eng}.} $k$. We have $\cl^\F(p,q)=\cl^\F_0(p,q)\oplus\cdots\oplus \cl^\F_n(p,q)$.

Clifford algebra $\cl^\F(p,q)$ is a superalgebra. It is represented as the direct sum of even and odd subspaces (of {\it even} and {\it odd} elements respectively)
\begin{eqnarray}
\cl^\F(p,q)=\cl^\F_{\Even}(p,q)\oplus\cl^\F_{\Odd}(p,q),\nonumber
\end{eqnarray}
$$\cl^\F_{\Even}(p,q)=\bigoplus_{k - even}\cl^\F_k(p,q),\qquad \cl^\F_{\Odd}(p,q)=\bigoplus_{k - odd}\cl^\F_k(p,q).$$

We introduce the operations of projection onto subspaces of rank-$k$ elements ($k=0, 1,\ldots, n$):
\begin{eqnarray}
\pi_k : \cl^\F(p,q)\to\cl_k^\F(p,q),\qquad \pi_k(U)=\sum_{a_1<\cdots<a_k}u_{a_1\ldots
a_k}e^{a_1\ldots a_k}.\label{rank:k:el}
\end{eqnarray}

The Clifford algebra $\cl^\F(p,q)$, $n=p+q$ has the following center
$${\rm Cen}(\cl^\F(p,q))=\left\lbrace\begin{array}{ll}
\cl^\F_0(p,q), & \mbox{if $n$ is even};\\
\cl^\F_0(p,q)\oplus \cl^\F_n(p,q) & \mbox{if $n$ is odd.}
\end{array}
\right.
$$

\noindent{\bf Pseudo-euclidian space $\R^{p,q}$ and changes of coordinates.} Let
$p,q$ be nonnegative integers and $n=p+q\geq1$. We denote an $n$-dimensional pseudo-Euclidian space of signature
$(p,q)$ with Cartesian coordinates $x^\mu$, $\mu=1,\ldots,n$ by
$\R^{p,q}$.
Tensor indices corresponding to the coordinates are denoted by small Greek letters. The metric tensor of pseudo-Euclidian space
$\R^{p,q}$ is given by the diagonal matrix of order $n$
\begin{equation}
\eta=\|\eta_{\mu\nu}\|=\|\eta^{\mu\nu}\|=\diag(1,\ldots,1,-1,\ldots,-1)\label{eta:matrix}
\end{equation}
with $p$ copies of $1$ and $q$ copies of $-1$ on the diagonal.

In $\R^{p,q}$ we deal with linear coordinate transformations\footnote{
We use the Einstein summation convention.  For example $p^\mu_\nu x^\nu=\sum_{\nu=1}^n p^\mu_\nu
x^\nu$.}
\begin{equation}
x^\mu\to\acute x^\mu=p^\mu_\nu x^\nu,\label{x:trans}
\end{equation}
preserving the metric tensor. So, real numbers $p^\mu_\nu$ satisfy relations
$p^\mu_\alpha p^\nu_\beta\eta^{\alpha\beta}=\eta^{\mu\nu}$, $p^\mu_\alpha p^\nu_\beta\eta_{\mu\nu}=\eta_{\alpha\beta}$.
In matrix formalism we can write $P^T\eta P=\eta$, $P\eta P^T=\eta$, where $T$ is the matrix transposition and the matrix
$P=\|p^\mu_\nu\|$ is from the pseudo-orthogonal group
$O(p,q)=\{ P\in\Mat(n,\R) : P^T\eta P=\eta\}.$

We denote the set of $(r,s)$ tensor fields (of rank $r+s$) of pseudo-Euclidian space $\R^{p,q}$ by $\T^r_s$. Real or complex tensor field $u\in\T^r_s$
has components $u^{\mu_1\ldots\mu_r}_{\nu_1\ldots\nu_s}$ in coordinates $x^\mu$. These components are smooth functions $\R^{p,q}\to\F$, where $\F$ is the field of real numbers $\R$ or complex numbers $\C$. In all considerations of this work it is sufficient that all functions of $x\in\R^{p,q}$ have continuous partial derivatives up to the second order.
\medskip

\noindent{\bf Functions with values in Clifford algebra.} Further we consider functions $\R^{p,q}\to\cl(p,q)$ with values in Clifford algebra. We assume that the basis elements (\ref{basis}) do not depend on the points $x\in\R^{p,q}$ i.e.
$$
\partial_\mu e^a=0,\quad \forall \mu, a=1, \ldots, n,
$$
where $\partial_\mu=\partial/\partial x^\mu$ are partial derivatives. The coefficients in the basis expansion of the Clifford algebra element $u_{a_1\ldots a_k}=u_{a_1\ldots
a_k}(x)$ may depend on $x\in\R^{p,q}$. In the present paper we also consider the functions with values in Lie algebras generated by the Clifford algebra (see p. \pageref{page:Lie}).

\medskip

\noindent{\bf Tensor fields with values in Clifford algebra.}
A tensor at the point $x\in\R^{p,q}$ with values in Clifford algebra is a mathematical object that belongs to the tensor product of the tensor algebra and Clifford algebra.

If a tensor field of rank $(r,s)$ in $\R^{p,q}$ has components $u^{\mu_1\ldots\mu_r}_{\nu_1\ldots\nu_s}=
u^{\mu_1\ldots\mu_r}_{\nu_1\ldots\nu_s}(x)$ in Cartesian coordinates $x^\mu$, then these components are considered as functions $\R^{p,q}\to\F$. These functions transform by the standard tensor transformation rule.

Components $U^{\mu_1\ldots\mu_r}_{\nu_1\ldots\nu_s}$ of tensor fields with values in Clifford algebra $\cl^\F(p,q)$ are considered as functions $\R^{p,q}\to\cl^\F(p,q)$ that transform under changes of coordinates by the standard tensor transformation rule.

We use the following notation for tensor fields with values in Clifford algebra: $U^{\mu_1\ldots\mu_r}_{\nu_1\ldots\nu_s}\in\cl(p,q)\T^r_s$ or $U\in\cl(p,q)\T^r_s$. In this notation the letter $\T$ means that this object is a tensor field. In particular, for scalar functions $U : \R^{p,q}\to\cl^\F(p,q)$
we use the notation $U\in\cl^\F(p,q)\T$.

For example, if we consider a tensor field $U^\mu_\nu\in\cl(p,q)\T^1_1$ with values in Clifford algebra, then we can write
\begin{eqnarray*}
U^\mu_\nu=u^\mu_\nu e+u^\mu_{\nu a} e^a+\sum_{a_1<a_2}u^\mu_{\nu a_1
a_2}e^{a_1 a_2}+\cdots+u^\mu_{\nu 1\ldots n}e^{1\ldots n},
\end{eqnarray*}
where $u^\mu_{\nu} , u^\mu_{\nu a}, u^\mu_{\nu a_1 a_2},\ldots,
u^\mu_{\nu1\ldots n}$ are real (in the case of $\cl^\R(p,q)$) or
complex (in the case of $\cl(p,q)$) tensor fields from $\T^1_1$.

In the present paper we also consider tensor fields with values in Lie algebras and scalar fields with values in Lie groups (see page \pageref{page:Lie}).

\medskip

\noindent{\bf Lie algebras in Clifford algebras.}

Let us consider the commutator (Lie bracket) $[U,V]=U V-V U$ of Clifford algebra elements $U,V\in\cl(p,q)$. This operation satisfy the Jacobi identity
$$
[[U,V],W] + [[V,W],U]+[[W,U],V]=0,\quad \forall U,V,W\in\cl(p,q).
$$
Therefore, Clifford algebra $\cl(p,q)$ can be considered as a Lie algebra with respect to the commutator. We can consider vector subspaces $L\subset\cl(p,q)$ of Clifford algebra that closed under commutator i.e. with the condition: if $U,V\in
L$ then $[U,V]\in L$. These subspaces are Lie algebras (generated by Clifford algebra), see also \cite{Doran}. Primarily we are interested in Lie algebras that are direct sums (as vector spaces) of subspaces of Clifford algebra elements of fixed ranks \cite{Shirokov}.

With the help of the operator $\pi_0 : \cl^\F(p,q)\to\cl_0^\F(p,q)$ we define operation of {\em Clifford algebra trace} $\Tr : \cl^\F\to\F$
$$
\Tr(U)=\pi_0(U)|_{e\to 1},\quad \forall U\in\cl^\F(p,q).
$$

\begin{theorem}{Theorem}\label{theoremTrPi} In Clifford algebra $\cl^\F(p,q)$ of arbitrary dimension $n=p+q$ we have
$\Tr([U,V])=0$ for all $U, V\in\cl^\F(p,q)$. In Clifford algebra $\cl^\F(p,q)$ of odd dimension $n=p+q$ we have $\pi_n([U ,V])=0$ for all $U, V\in\cl^\F(p,q)$.
\end{theorem}

\proof. Using (\ref{cond}) for 2 arbitrary basis elements we obtain
\begin{eqnarray}
[e^{a_1\ldots a_k}, e^{b_1 \ldots b_l}]=(1-(-1)^{kl-s})e^{a_1\ldots a_k}e^{b_1 \ldots b_l}\in\cl^\F_{k+l-2s}(p,q),\label{dok}
\end{eqnarray}
where $s$ is the number of coincident indices in the ordered multi-indices $a_1 \ldots a_k$ and $b_1 \ldots b_l$. If $k=l=s$, then $1-(-1)^{kl-s}$ equals $0$. If $k+l=n$, $s=0$ and $n$ is odd, then it equals $0$ again. For more details see Theorem 1 in \cite{Shirokov}.
\rule{5pt}{5pt}

Consider the set of Clifford algebra elements with zero projection onto Clifford algebra center
$$\cl_\circledS(p,q)=\cl(p,q)\setminus {\rm Cen}(\cl(p,q)).$$

\begin{theorem}{Theorem}
The set $\cl_\circledS(p,q)$ is a Lie algebra with respect to the commutator $[A,B]=AB-BA$.
\end{theorem}

\proof. See the previous theorem. \rule{5pt}{5pt}

\begin{theorem}{Theorem}\label{LemmaDif}
Let $F=F(x)$ be a function with values in the Lie algebra $\cl_\circledS(p,q)$. Then the partial derivatives
$\partial_\mu F$ are functions (components of a covariant vector field) with values in the same Lie algebra $\cl_\circledS(p,q)$.
\end{theorem}

\proof. If $n$ is even, then the function $F=F(x)$ can be written as basis expansion (\ref{basis})
\begin{equation}
F=f_a e^a+\sum_{a_1<a_2}f_{a_1a_2}e^{a_1a_2}+\cdots+f_{1\ldots n}e^{1\ldots n},\label{F:decomp}
\end{equation}
Since $\Tr\,F=0$, then the first term $f e$ is absent. We assume that Clifford algebra generators $e^a$ do not depend on $x\in\R^{p,q}$.  So
$\partial_\mu e^a=0$ for all $\mu,a=1,\ldots n$ and
$$
\partial_\mu F=(\partial_\mu f_a)e^a+\sum_{a_1<a_2}(\partial_\mu f_{a_1a_2})e^{a_1a_2}+\cdots+
(\partial_\mu f_{1\ldots n})e^{1\ldots n}.
$$
We obtain $\Tr\,F=0$ and $\Tr(\partial_\mu F)=0$, i.e. $\partial_\mu F\in\cl_\circledS(p,q)$.

If $n$ is odd, then the function $F=F(x)\in\cl_\circledS(p,q)$ can be written as basis expansion (\ref{F:decomp}) without the first term $f e$ and without the last term $f_{1\ldots n}e^{1\ldots n}$. We obtain $\partial_\mu F\in\cl_\circledS(p,q)$ again.
\rule{5pt}{5pt}

The following subspaces of Clifford algebra are Lie algebras with respect to the commutator:
$\cl_2(p,q)$, $\cl_1(p,q)\oplus\cl_2(p,q)$, $\cl_2(p,q)\oplus\cl_3(p,q)$, $\cl_0(p,q)$, ${\rm Cen}(\cl(p,q))$, $\cl_\circledS(p,q)$.

In the Section 2 we have considered gamma-matrices in Dirac representation (which are used in the Dirac equation for an electron) and found that $i\gamma^\mu\in\su(2,2)$. In Clifford algebra
$\cl(p,q)$ the following Lie algebra is analogue\footnote{See \cite{mybook:eng, MarShir:book}.} of the Lie algebra $\su(2,2)$:
$$
\w(\cl(p,q)) =\bigoplus^{\acute n}_{k=1}i^{\frac{k(k-1)}{2}+1}\cl^\R_k(p,q)
$$
where $\acute n=n$ in the case of even $n$ and $\acute n=n-1$ in the case of odd $n$.
%In other words:
%$$
%\w(\cl(p,q))
%=i\cl^\R_1(p,q)\oplus\cl^\R_2(p,q)\oplus\cl^\R_3(p,q)\oplus
%i\cl^\R_4(p,q)\oplus\cdots\oplus a_{\acute n}\cl^\R_{\acute
%n}(p,q),
%$$
%where $a_k=i$ if $k=0,1\ \mod\ 4$ and $a_k=1$ if $k=2,3\ \mod\ 4$.

We are interested in Lie subalgebras of this Lie algebra. As we will see, Lie algebra $\cl^\R_2(p,q)$ plays an important role in field theory equations. Other important Lie algebras contain Lie subalgebra
$i\cl^\R_1(p,q)\oplus\cl^\R_2(p,q)$:\label{page:Lie}
\begin{itemize}
\item For $n\geq2$: $i\cl^\R_1(p,q)\oplus\cl^\R_2(p,q)$.
\item For $n\geq6$: $i\cl^\R_1(p,q)\oplus\cl^\R_2(p,q)\oplus a_{\acute n-1}\cl^\R_{\acute n-1}(p,q)\oplus a_{\acute n}\cl^\R_{\acute n}(p,q),$ where $\acute n=n$ for even $n$ and $\acute n=n-1$ for odd $n$.
\item For $n\geq8$: $i\cl^\R_1(p,q)\oplus\cl^\R_2(p,q)\oplus i\cl^\R_5(p,q)\oplus\cl^\R_6(p,q)\oplus i\cl^\R_9(p,q)\oplus\cl^\R_{10}(p,q)\oplus\cdots\oplus a_r\cl^\R_r(p,q)$, where
$r=n-2$ if $n=0\mod 4$, $r=n-3$ if $n=1\mod 4$, $r=n$ if $n=2\mod 4$, $r=n-1$ if $n=3\mod 4$.
\end{itemize}

We consider pinor groups as the following sets of Clifford algebra elements:
$$
\Pin(p,q) =
\{S\in\cl_\Even^\R(p,q)\,\mbox{or}\,S\in\cl^\R_\Odd(p,q) : S^\sim
S=\pm e,\, S^{-1}e^a S\in\cl_1^\R(p,q)\},
$$
where linear operation $\sim : \cl_k(p,q)\to\cl_k(p,q)$,
$k=0,1,\ldots,n$ is called {\em reversion}. This operation reverses the order of generators in products:
$(e^{a_1}\cdots e^{a_k})^\sim=e^{a_k}\cdots e^{a_1}$.

Note that the set of rank 2 Clifford algebra elements $\cl_2^\F(p,q)$ is closed w.r.t. commutator and hence generates a Lie algebra. The Lie algebra $\cl_2^\R(p,q)\subset \w(\cl(p,q))$
is a real Lie algebra of the Lie group $\Pin(p,q)$ (see
\cite{p-adic}).

\section{Relation between projection operators and contractions in Clifford algebras}\label{section:projector}

Consider operations of projection (\ref{rank:k:el}) onto subspaces $\cl_k(p,q)$ of Clifford algebra elements of rank $k$.

The following sum is called a {\em generator contraction} of an arbitrary Clifford algebra element $U\in\cl(p,q)$:
\begin{equation}
F(U)=e^a Ue_a,\label{svert}
\end{equation}
where $e_a=\eta_{ab}e^b$. We use notations $F^0(U)=U$, $F^1(U)=F(U)$, $F^2(U)=F(F(U))$, etc.
Note that $F^l: \cl_k(p,q)\to \cl_k(p,q)$ for all $k, l=0, 1, \ldots, n$.

According to the theorem on generator contraction \cite{MarSh2008} we have
\begin{equation}
F(U)=\sum_{k=0}^n \lambda_k \pi_k(U),\qquad \mbox{where}\quad\lambda_k=(-1)^k(n-2k).\label{svert2}
\end{equation}

\begin{theorem}{Theorem} Consider an arbitrary Clifford algebra element $U\in\cl(p,q)$, $n=p+q$. Then we have
\begin{eqnarray}
\pi_k(U)=\sum_{l=0}^n b_{kl} F^l(U)\quad \mbox{if $n$ is even;}\qquad \pi_{k, n-k}(U)=\sum_{l=0}^{\frac{n-1}{2}}g_{kl} F^l(U)\quad \mbox{if $n$ is odd,\label{svpr2}}
\end{eqnarray}
where $B=||b_{kl}||$ is inverse of matrix $A_{(n+1) \times (n+1)}=||a_{kl}||$, $a_{kl}=(\lambda_{l-1})^{k-1}$, $G=||g_{kl}||$ is inverse of matrix $D_{\frac{n+1}{2} \times \frac{n+1}{2}}=||d_{kl}||$, $d_{kl}=(\lambda_{l-1})^{k-1}$ and $\lambda_k=(-1)^k(n-2k)$ and
$\pi_{k, n-k}=\pi_k+\pi_{n-k}$ is operation of projection onto subspace $\cl_k(p,q)\oplus\cl_{n-k}(p,q)$.
\end{theorem}

\proof. We have $F^l(U)=\sum_{k=0}^n (\lambda_k)^l \pi_k(U),$
then
$$
\left( \begin{array}{l}
 F^0(U) \\
 F^1(U) \\
 \ldots \\
 F^n(U) \end{array}\right)=
 A
 \left( \begin{array}{l}
 \pi_0(U) \\
 \pi_1(U) \\
 \ldots \\
 \pi_n(U) \end{array}\right),\qquad A=\left( \begin{array}{llll}
 1 & 1 & \ldots & 1\\
 \lambda_0 & \lambda_1 & \ldots & \lambda_n\\
 \ldots & \ldots & \ldots & \ldots\\
 (\lambda_0)^n & (\lambda_1)^n & \ldots & (\lambda_n)^n \end{array}\right).
$$

Matrix $A$ is a Vandermonde matrix. Its determinant equals
$$\det A=\prod_{0\leq i < j \leq n}(\lambda_j -\lambda_i).$$

In the case of even $n$ we have $\lambda_k=-\lambda_{n-k}$, because $\lambda_{n-k}=(-1)^{n-k}(n-2(n-k))=(-1)^k (2k-n)=-\lambda_{k}$.
In particular, $\lambda_{\frac{n}{2}}=0.$
It is easy to see that all $\lambda_k$ are different in the case of even $n$, and Vandermonde matrix is invertible. Denote the inverse matrix by $B=||b_{ij}||$:
$$ \left( \begin{array}{l}
 \pi_0(U) \\
 \pi_1(U) \\
 \ldots \\
 \pi_n(U) \end{array}\right)=
 \left( \begin{array}{llll}
 b_{00} & b_{01} & \ldots & b_{0n}\\
 b_{10} & b_{11} & \ldots & b_{1n}\\
 \ldots & \ldots & \ldots & \ldots\\
 b_{n0} & b_{n1} & \ldots & b_{nn} \end{array}\right)
 \left( \begin{array}{l}
 F^0(U) \\
 F^1(U) \\
 \ldots \\
 F^n(U) \end{array}\right).
$$
There exists the explicit formula for inverse of Vandermonde matrix but we do not use it.

In the case of odd $n$ we have $\lambda_k=\lambda_{n-k}$, and hence Vandermonde matrix is singular and projection operations do not expressed through contractions. In this case we use projections $\pi_{k, n-k}$:
$$
\left( \begin{array}{l}
 F^0(U) \\
 F^1(U) \\
 \ldots \\
 F^{\frac{n-1}{2}}(U) \end{array}\right)=
 \left( \begin{array}{llll}
 1 & 1 & \ldots & 1\\
 \lambda_0 & \lambda_1 & \ldots & \lambda_{\frac{n-1}{2}}\\
 \ldots & \ldots & \ldots & \ldots\\
 (\lambda_0)^{\frac{n-1}{2}} & (\lambda_1)^{\frac{n-1}{2}} & \ldots & (\lambda_{\frac{n-1}{2}})^{\frac{n-1}{2}} \end{array}\right)
 \left( \begin{array}{l}
 \pi_{0, n}(U) \\
 \pi_{1, n-1}(U) \\
 \ldots \\
 \pi_{\frac{n-1}{2}, \frac{n+1}{2}}(U) \end{array}\right).
$$

We denote the invertible matrix from the last formula by $D$ and inverse of $D$ by $G=||g_{ij}||$.

We obtain the relation between projection operations and contractions in the following form:
$$ \left( \begin{array}{l}
 \pi_{0, n}(U) \\
 \pi_{1, n-1}(U) \\
 \ldots \\
 \pi_{\frac{n-1}{2}, \frac{n+1}{2}}(U) \end{array}\right)=
 \left( \begin{array}{llll}
 g_{00} & g_{01} & \ldots & g_{0 \frac{n-1}{2}}\\
 g_{10} & g_{11} & \ldots & g_{1 \frac{n-1}{2}}\\
 \ldots & \ldots & \ldots & \ldots\\
 g_{\frac{n-1}{2} 0} & g_{\frac{n-1}{2} 1} & \ldots & g_{\frac{n-1}{2} \frac{n-1}{2}} \end{array}\right)
 \left( \begin{array}{l}
 F^0(U) \\
 F^1(U) \\
 \ldots \\
 F^\frac{n-1}{2}(U) \end{array}\right).\quad \rule{5pt}{5pt}
$$

So, in the case of even $n$ operations of projection of Clifford algebra elements $U\in\cl(p,q)$ is uniquely expressed through contractions (of order not more than $n$) of element $U$. Note that we can use these formulas as the definition of operations of projection onto subspaces of fixed ranks.

Let's give some examples. In the case of $n=2$ we have
\begin{eqnarray}
&&A=\left( \begin{array}{ccc}
 1 & 1 & 1 \\
 2 & 0 & -2 \\
 4 & 0 & 4 \end{array}\right),\qquad
 B=\left( \begin{array}{ccc}
 0 & \frac{1}{4} & \frac{1}{8} \\
 1 & 0 & -\frac{1}{4} \\
 0 & -\frac{1}{4} & \frac{1}{8} \end{array}\right),\nonumber\\
&&F^0(U)=U,\quad F^1(U)= 2\pi_0(U)-2\pi_2(U),\quad F^2(U)=4\pi_0(U)+4\pi_2(U),\nonumber\\
&&\pi_0(U)=\frac{1}{4}e^a U e_a +\frac{1}{8}e^a e^b U e_b e_a,\qquad \pi_1(U)=U-\frac{1}{4}e^a e^b U e_a e_b,\nonumber\\
&&\pi_2(U)=-\frac{1}{4}e^a U e_a +\frac{1}{8}e^a e^b U e_b e_a.\nonumber
\end{eqnarray}
In the case of $n=4$ we have
\begin{equation}
A=\left( \begin{array}{ccccc}
 1 & 1 & 1 & 1 & 1\\
 4 & -2 & 0 & 2 & -4\\
 16 & 4 & 0 & 4 & 16\\
 64 & -8 & 0 & 8 & -64\\
 256 & 16 & 0 & 16 & 256 \end{array}\right),\quad
B=\left( \begin{array}{ccccc}
 0 & -\frac{1}{24}
      & -\frac{1}{96}  & \frac{1}
   {96} & \frac{1}{384} \cr
   0 & -\frac{1}
     {3} & \frac{1}{6} & \frac{1}
   {48} & -\frac{1}{96}
      \cr 1 & 0 & -\frac{5}{16}
      & 0 & \frac{1}{64} \cr 0 & \frac{1}
   {3} & \frac{1}{6} & -\frac{1}{48}
      & -\frac{1}{96}
      \cr 0 & \frac{1}{24} & -\frac{1}
     {96}  & -\frac{1}{96}
       & \frac{1}{384} \cr   \end{array}\right).
      \nonumber
\end{equation}
In the case of odd dimension $n=3$ we have
\begin{eqnarray}
F^0(U)&=&U=\pi_0(U)+\pi_1(U)+\pi_2(U)+\pi_3(U),\nonumber\\
F^1(U)&=& 3\pi_0(U)-\pi_1(U)-\pi_2(U)+3\pi_3(U),\nonumber\\
F^2(U)&=&9\pi_0(U)+\pi_1(U)+\pi_2(U)+9\pi_3(U),\nonumber\\
F^3(U)&=&27\pi_0(U)-\pi_1(U)-\pi_2(U)+27\pi_3(U).\nonumber
\end{eqnarray}

Matrix of this system of equations is singular. But we can consider expressions
$\pi_{03}(U)=\pi_0(U)+\pi_3(U)$, $\pi_{12}(U)=\pi_1(U)+\pi_2(U)$
and obtain
\begin{eqnarray}
F^0(U)&=&U=\pi_{03}(U)+\pi_{12}(U),\qquad F^1(U)=3\pi_{03}(U)-\pi_{12}(U),\nonumber\\
\pi_{03}(U)&=&\frac{1}{4}F^0(U)+\frac{1}{4}F^1(U)=\frac{1}{4}U +\frac{1}{4}e^a U e_a,\nonumber\\
\pi_{12}(U)&=&\frac{3}{4}F^0(U)-\frac{1}{4}F^1(U)=\frac{3}{4}U-\frac{1}{4}e^a  U e_a,\nonumber\\
D&=&\left( \begin{array}{cc}
 1 & 1  \\
 3 & -1 \end{array}\right),\qquad
 G=\left( \begin{array}{cc}
 \frac{1}{4} & \frac{1}{4}  \\
 \frac{3}{4} & -\frac{1}{4} \end{array}\right).\nonumber
\end{eqnarray}

\section{Clifford field vectors and an algebra of $h$-forms.}

In this section we introduce new geometric objects - Clifford field vector and an algebra of h-forms which is a generalization of the algebra of differential forms and the Atiyah-K\"{a}hler algebra \cite{atiyah, kahler}. We combine the technique of the Dirac gamma matrices and the technique of differential forms, in particular, the Atiyah-K\"{a}hler algebra of differential forms. From our point of view, these objects are helpful for consideration of some problems related to field theory equations.

\noindent{\bf Frame field $y^\mu_a$.} A set of $n$ real vector fields $y^\mu_a=y^\mu_a(x)\in\T^1$ of pseudo-Euclidian space $\R^{p,q}$ enumerated by the Latin index ($a=1,\ldots,n$) and satisfying
$$
y^\mu_a y^\nu_b\eta^{ab}=\eta^{\mu\nu},\quad \forall x\in\R^{p,q}
$$
is called  a {\it frame field}. Using local (that depends on $x$) pseudo-orthogonal transformation, we can get another frame field from the frame field $y^\mu_a$
$$
y^\mu_a\to\hat y^\mu_a=q^b_a y^\mu_b,
$$
where $q_a^b=q_a^b(x)$ are smooth functions of $x\in\R^{p,q}$ and matrix $Q=Q(x)=\|q^b_a\|$ is such that $Q\in O(p,q)$ for any $x$. It is easy to see that
$$
\hat y^\mu_a\hat y^\nu_b\eta^{ab}=\eta^{\mu\nu},\quad \forall
x\in\R^{p,q}
$$
i.e. the set of $n$ vector fields $\hat y^\mu_a$ is also a frame field.
\medskip

\noindent{\bf Coframe field $y^b_\nu$.} A set of $n$ real covector fields $y^b_\nu=y^b_\nu(x)\in\T_1$ of pseudo-Euclidian space $\R^{p,q}$ enumerated by the Latin index
($b=1,\ldots,n$) and satisfying
$$
y_\mu^a y_\nu^b\eta_{ab}=\eta_{\mu\nu},\quad \forall x\in\R^{p,q}
$$
is called a {\it coframe field}.

If we have frame field $y^\mu_a$, then we can get coframe field using Minkowski matrix:
$$
y^b_\nu=\eta^{ab}\eta_{\mu\nu}y_a^\mu.
$$

\medskip

\noindent{\bf  Clifford field vector $h^\mu$.}  If
$h^\mu=h^\mu(x)$ are components of vector field with values in $\cl(p,q)$ that satisfy
the following relations:
 \begin{equation}
h^\mu h^\nu+h^\nu h^\mu=2\eta^{\mu\nu}e, \quad \mu,\nu=1,\ldots,n  \label{main:id:h}
\end{equation}
for any $\forall x\in\R^{p,q}$ and the condition
\begin{equation}
\Tr(h^1\cdots h^n)=0, \label{h1hn}
\end{equation}
then the vector $h^\mu\in\cl(p,q)\T^1$ is called a {\em Clifford field vector}\footnote{In this expression the meaning of the word ``field'' is the same as in ``field theory'' (not as in ``tensor field'')}.

Note that condition (\ref{h1hn}) holds automatically in the case of even $n$, i.e. this condition is necessary for the case of odd $n$.

Denote the set of invertible Clifford algebra elements by $\cl^\times(p,q)$. Note that $\cl^\times(p,q)$ is a Lie group with respect to the Clifford multiplication.

If $h^\mu$ is a Clifford field vector and $S\in\cl^\times(p,q)\T$ is continuous function, then we can get the pair of new Clifford field vectors using similarity transformation \footnote{In the case of even $n$ it is sufficiently to consider only relation $\hat h^\mu=S^{-1}h^\mu S$ (see \cite{MarShir:book}).} $\hat h^\mu=\pm S^{-1}h^\mu S.$

For example, let us consider a frame field $y^\mu_a=y^\mu_a(x)$ and a smooth function $S\in\cl^\times(p,q)\T$ with values in the set of invertible Clifford algebra elements. With the help of generators $e^a$ we get the vector field
$$
h^\mu=h^\mu(x)=y^\mu_a S^{-1}e^a S\in\cl(p,q)\T^1.
$$
It is easy to see that components of this vector field satisfy relations (\ref{main:id:h}) and (\ref{h1hn}), i.e. $h^\mu$ is a Clifford field vector.

Components of field vector transform under (orthogonal) changes of coordinates (\ref{x:trans}) using standard tensor transformation law
\begin{equation}
h^\mu\to\acute h^\mu=p^\mu_\nu h^\nu,\qquad P=\|p^\mu_\nu\|\in{\rm O}(p,q).\label{h:trans}
\end{equation}

With the help of the metric tensor we can raise and lower indices:
$$
h_\nu=\eta_{\mu\nu}h^\mu,\quad h^\mu=\eta^{\mu\nu}h_\nu.
$$

\begin{theorem}{Theorem}\label{theorem:T1}
If $n=p+q\geq2$ and $h^\mu$ is a Clifford field vector, then $h^\mu\in\cl_\circledS(p,q)\T^1$.
\end{theorem}

\proof. Let us consider a coframe field $y_\mu^a$. We define $n$ elements
$h^a=y^a_\mu h^\mu\in\cl(p,q)$, satisfying
$h^a h^b+h^b h^a=2\eta^{ab}e$ for all $a,b=1,\ldots,n$.

Let $n=p+q$ be even. We prove that for any $x\in\R^{p,q}$ we have $\Tr\,h^\mu=0$. By the generalized Pauli's theorem \cite{OTP} there exists an invertible element $S\in\cl(p,q)$ (at any $x\in\R^{p,q}$) such that
$h^a=S^{-1}e^a S$, $a=1,\ldots n$. So
$$
\Tr\,h^a=\Tr(S^{-1}e^a S)=\Tr\,e^a=0,\quad \Tr\,h^\mu=\Tr(y^\mu_a
h^a)=0.
$$
It proves the theorem for the case of even $n$.

Let $n=p+q\geq 3$ be odd. We prove that for any $x\in\R^{p,q}$ we have $\Tr\,h^\mu=0$ and $\Tr(e^{1\ldots n}h^\mu)=0$.
By the generalized Pauli's theorem \cite{OTP} there exists an invertible element $S\in\cl(p,q)$ (at any $x\in\R^{p,q}$)
such that two sets of $n$ elements $\{e^a\}$ and $\{h^a\}$ are related by one of two following formulas:
$h^a=\epsilon\, S^{-1}e^a S$,  $a=1,\ldots n$, $\epsilon=\pm1$. Then
$e^{1\ldots n}h^a=\epsilon\,e^{1\ldots n} S^{-1}e^a S$, $a=1,\ldots, n$.

Note that the element $e^{1\ldots n}e^a$ is an element of rank $n-1$. Therefore $\Tr(e^{1\ldots n}e^a)=0$ for $n>1$. Since the element $e^{1\ldots n}$ ($n$ - odd) is from the center of Clifford algebra $\cl(p,q)$, then for $n\geq3$
$$
\Tr\,h^a=\epsilon\,\Tr\,e^a=0,\quad\Tr(e^{1\ldots
n}h^a)=\epsilon\,\Tr(e^{1\ldots n}e^a)=0,\quad a=1,\ldots, n.
$$
Consequently, for odd $n\geq3$ and for any $x\in\R^{p,q}$ we have
$$
\Tr\,h^\mu=0,\quad\Tr(e^{1\ldots n}h^\mu)=0,\quad a=1,\ldots, n.
$$
It means that $h^\mu\in\cl_\circledS(p,q)\T^1$. \rule{5pt}{5pt}

\noindent{\bf $h$-forms.} Let us consider a covariant skew-symmetric tensor field
$u_{\mu_1\ldots\mu_k}\in\T_{[k]}$ of rank $k$ and a Clifford field vector $h^\mu\in\cl(p,q)\T^1$. We say that the expression
$$
\frac{1}{k!}u_{\mu_1\ldots\mu_k}h^{\mu_1}\cdots
h^{\mu_k}=\sum_{\nu_1<\cdots<\nu_k}u_{\nu_1\ldots\nu_k}h^{\nu_1}\cdots
h^{\nu_k}
$$
is an {\em $h$-form of rank $k$}.

If we have a scalar function $u=u(x)$ and $n$ covariant skew-symmetric tensor fields
$u_{\mu_1\ldots\mu_k}\in\T_{[k]}$ of ranks $k=1, 2,\ldots, n$, then we say that
\begin{equation}
U= u e+\sum_{k=1}^n\frac{1}{k!}u_{\mu_1\ldots\mu_k}h^{\mu_1}\cdots
h^{\mu_k}=u
e+\sum_{k=1}^n\sum_{\nu_1<\cdots<\nu_k}u_{\nu_1\ldots\nu_k}h^{\nu_1}\cdots
h^{\nu_k} \label{h:form}
\end{equation}
is an {\em $h$-form} or a {\em heterogeneous $h$-form}.

An $h$-form is invariant under orthogonal changes of coordinates (\ref{x:trans}). Components $u_{\mu_1\ldots\mu_k}$ of an $h$-form are components of covariant skew-symmetric tensor fields of ranks $k=0,\ldots,n$.

If we do not pay attention to the difference between tensor (Greek) and nontensor (Latin) indices\footnote{The difference between tensor and nontensor indices appears only when we consider coordinate transformations of pseudo-Euclidian space $\R^{p,q}$.}, then, by relations (\ref{main:id:h}), we can consider components of the field vector $h^\mu$ as generators of Clifford algebra. A set of $h$-forms over the field $\F$ is called the {\em algebra of $h$-forms} $\cl[h]^\F(p,q)$\footnote{In notation $\cl[h]^\F(p,q)$ symbol $h$ means that the basis is generated by Clifford field vector $h^\mu$.}. We denote the set of $h$-forms of rank $k$ by $\cl[h]^\F_k(p,q)$. If $U$ is an $h$-form (\ref{h:form}) then
we denote projections of $U$ onto $\cl[h]^\F_k(p,q)$ by
$\pi[h]_k(U)$, $k=0,1,\ldots,n$. To calculate projections $\pi[h]_k(U)$ we can use method of contractions by components of Clifford field vector using Vandermonde matrix (as in the section \ref{section:projector}). Structure of algebra of $h$-forms is considered as a geometrization of structure of Clifford algebra.

Lie algebras generated by Clifford algebra were considered on the page \pageref{page:Lie}. We will use the following Lie algebras generated by the algebra of $h$-form:\label{page:Lie:h}
\begin{eqnarray*}
&&\cl[h]_2(p,q),\ \cl[h]_1(p,q)\oplus\cl[h]_2(p,q),\ \cl[h]_2(p,q)\oplus\cl[h]_3(p,q),\\
&& {\rm Cen}(\cl[h](p,q)),\ \cl[h]_\circledS(p,q),
\end{eqnarray*}
where ${\rm Cen}(\cl[h](p,q))$ is the center of algebra of $h$-forms,
$\cl[h]_\circledS(p,q)=\cl[h](p,q)\setminus{\rm Cen}(\cl[h](p,q))$ is the set of $h$-forms with zero projection onto the center of algebra of $h$-forms.

Note that $\cl[h]_\circledS(p,q)\simeq\cl_\circledS(p,q)$, because $\cl[h]_0(p,q)\simeq\cl_0(p,q)$ for any natural $n=p+q$ and $\cl[h]_n(p,q)\simeq\cl_n(p,q)$ for any odd $n$.

\medskip

\noindent{\bf  Tensor fields with values in $h$-forms.}
Tensor field $U^{\nu_1\ldots\nu_k}_{\rho_1\ldots\rho_r}$ with values in $h$-forms (at point $x\in\R^{p,q}$) belongs to the tensor product of tensor algebra and the algebra of $h$-forms. We write $U^{\nu_1\ldots\nu_k}_{\rho_1\ldots\rho_r}\in\cl[h](p,q)\T^k_r$. For example, tensor field $U^\nu_\rho\in\cl[h](p,q)\T^1_1$ can be represented as
$$ U^\nu_\rho=u^\nu_\rho
e+\sum_{k=1}^n\frac{1}{k!}u^\nu_{\rho\mu_1\ldots\mu_k}h^{\mu_1}\cdots
h^{\mu_k},
$$
where $u^\nu_{\rho\mu_1\ldots\mu_k}=u^\nu_{\rho[\mu_1\ldots\mu_k]}$
are components of $(1,k+1)$ tensor field which are skew-symmetric w.r.t. $k$ covariant indices (antisymmetrization is denoted by square brackets).

Note that we can consider Clifford field vector $h^\mu$ as vector with values in $h$-forms of rank $1$. Actually,
$h^\mu=\delta^\mu_\nu h^\nu\in\cl[h]_1(p,q)\T^1$, where $\delta^\mu_\nu$ is Kronecker tensor ($\delta^k_r=0$ if
$k\neq r$ and $\delta^k_r=1$ if $k=r$). Also we have $h_\mu=\eta_{\mu\nu} h^\nu\in\cl[h]_1(p,q)\T_1$, where $\eta_{\mu\nu}$ are components of metric tensor of pseudo-Euclidian space $\R^{p,q}$.

Note that we also consider tensor fields with values in Lie algebras generated by algebra of $h$-form in this paper (see p. \pageref{page:Lie:h})).

\section{Primitive field equation and its gauge symmetry}

Consider the equation (system of equations)
\begin{equation}
\partial_\mu h_\rho-[C_\mu, h_\rho]=0,\quad \mu,\rho=1,\ldots,n,\label{nik:eq}
\end{equation}
where $h^\rho\in\cl(p,q)\T^1$ is an arbitrary Clifford field vector and $C_\mu=C_\mu(x)$ ($x\in\R^{p,q}$) is covector field with values in $\cl(p,q)$.

We consider system of equations (\ref{nik:eq}) as a new field equation. This equation is called {\em a primitive field equation}.

Note that if we have a solution $C_\mu=C_\mu(x)\in\cl(p,q)\T_1$ of system of equations (\ref{nik:eq}) and $\alpha_\mu=\alpha_\mu(x)$ are arbitrary continuous components of covector field with values in center of Clifford algebra, then components $C_\mu+\alpha_\mu\in\cl(p,q)\T_1$ also satisfy equation (\ref{nik:eq}).

Therefore it is reasonable to assume that $C_\mu\in\cl_\circledS(p,q)\T_1$.

\begin{theorem}{Theorem}
Let $h^\nu\in\cl_\circledS(p,q)\T^1$ be a Clifford field vector and $C_\mu\in\cl_\circledS(p,q)\T_1$ satisfy the primitive field equation
$$
\partial_\mu h_\rho-[C_\mu,h_\rho]=0,\quad \forall \mu,\rho=1, \ldots, n.
$$
Let $S : \R^{p,q}\to\cl^\times(p,q)$ be a function with values in $\cl^\times(p,q)$ such that
$S^{-1}\partial_\mu S\in\cl_\circledS(p,q)\T_1.$ Then, the following components of covectors
$$
\acute h_\rho=S^{-1}h_\rho S\in\cl_\circledS(p,q)\T_1,\quad
\acute C_\mu=S^{-1}C_\mu S-S^{-1}\partial_\mu S\in\cl_\circledS(p,q)\T_1
$$
also satisfy the equation
$\partial_\mu \acute h_\rho-[\acute C_\mu,\acute h_\rho]=0,\quad \forall \mu,\rho=1,\ldots,n.$
\end{theorem}

\proof. The condition $\acute h_\rho\in\cl_\circledS(p,q)\T_1$ holds automatically for every $S\in\cl^\times(p,q)T$ because $\Tr(S^{-1}h_\rho S)=\Tr(h_\rho)$ in the case of natural $n$ and $\pi[h]_n(S^{-1}h_\rho S)=\pi[h]_n(h_\rho)$ in the case of odd $n$ (see Theorems \ref{theoremTrPi} and \ref{theorem:T1}).

To satisfy the condition $\acute C_\mu\in\cl_\circledS(p,q)\T_1$ we need functions $S$ from the class ${\mathbb S}$
$${\mathbb S}=\{S\in\cl^\times(p,q)T : S^{-1}\partial_\mu S\in\cl_\circledS(p,q)T_1\}.$$

Then
\begin{eqnarray}
&&\partial_\mu \acute h_\rho-[\acute C_\mu, \acute h_\rho]=\partial_\mu(S^{-1}h_\rho S)-(S^{-1}C_\mu S-S^{-1}\partial_\mu S)S^{-1}h_\rho S\nonumber\\
&&+S^{-1}h_\rho S(S^{-1}C_\mu S-S^{-1}\partial_\mu S)=\partial_\mu S^{-1}h_\rho S+S^{-1}\partial_\mu h_\rho S+S^{-1}h_\rho \partial_\mu S\nonumber\\
&&-S^{-1}C_\mu h_\rho S+S^{-1}\partial_\mu S S^{-1}h_\rho S+S^{-1}h_\rho C_\mu S-S^{-1}h_\rho \partial_\mu S\nonumber\\
&&=S^{-1}(\partial_\mu h_\rho-[C_\mu, h_\rho])S+S^{-1}(S\partial_\mu S^{-1}+\partial_\mu S S^{-1})h_\rho S=0.\nonumber \quad \rule{5pt}{5pt}
\end{eqnarray}

Remark. Professor G.~A.~Alekseev called our attention to the following fact. If we consider elements $S=S(x)$ as matrices then we can use the well known formula
$$
\Tr(S^{-1}\partial_\mu S)=\partial_\mu(\ln(\det\,S)).
$$
By this formula, from the condition $S^{-1}\partial_\mu
S\in\cl_\circledS(p,q)\T_1$ it follows that $\det\,S$ does not depend on $x\in\R^{p,q}$. So we may normalize $S$ and take $\det S=1$ or $\det S=-1$.
\medskip

\begin{theorem}{Theorem}\label{theorem:CmuCnu}
Let $h^\mu\in\cl_\circledS(p,q)\T^1$ be a Clifford field vector and $C_\mu\in\cl_\circledS(p,q)\T_1$ be a covector field. If $h^\mu$ and $C_\nu$ are related by equation
$$
\partial_\mu h^\nu-[C_\mu,h^\nu]=0,\qquad \forall \mu,\nu=1,\ldots,n,
$$
then components of covector field $C_\mu$ satisfy the conditions
\begin{equation}
\partial_\mu C_\nu-\partial_\nu C_\mu-[C_\mu,C_\nu]=0,\qquad \forall\mu,\nu=1,\ldots,n.\label{CmuCnu}
\end{equation}
Conditions (\ref{CmuCnu}) are invariant under the gauge transformation
$$
C_\mu\to\acute C_\mu=S^{-1}C_\mu S-S^{-1}\partial_\mu S,
$$
where $S=S(x)$ is a function from ${\mathbb S}$, i.e. $S\in\cl^\times(p,q)T$ and $S^{-1}\partial_\mu S\in\cl_\circledS(p,q)\T_1$.
\end{theorem}

\proof. Let us differentiate conditions $\partial_\mu h^\lambda=[C_\mu,h^\lambda],$
and obtain
\begin{eqnarray}
\partial_\nu\partial_\mu h^\lambda&=&[\partial_\nu C_\mu,h^\lambda]+[C_\mu,\partial_\nu h^\lambda]=[\partial_\nu C_\mu,h^\lambda]+[C_\mu,[C_\nu,h^\lambda]],\nonumber\\
0&=&(\partial_\mu\partial_\nu-\partial_\nu\partial_\mu)h^\lambda=[\partial_\mu C_\nu-\partial_\nu C_\mu-[C_\mu,C_\nu],h^\lambda].\label{dd-dd:new}
\end{eqnarray}
If an element of Clifford algebra commutes with all generators (with $h^\mu$, $\mu=1,\ldots,n$ in this case), then this element belongs to the center of Clifford algebra. Therefore, from (\ref{dd-dd:new}) implies
\begin{eqnarray}
\partial_\mu C_\nu-\partial_\nu C_\mu-[C_\mu,C_\nu]&=&c_{\mu\nu}e,\quad \mbox{if $n=p+q$ is even},\nonumber\\
\partial_\mu C_\nu-\partial_\nu C_\mu-[C_\mu,C_\nu]&=&c_{\mu\nu}e+d_{\mu\nu}e^1\cdots e^n,\quad \mbox{if $n=p+q$ is odd},\nonumber
\end{eqnarray}
where $c_{\mu\nu}$, $d_{\mu\nu}$ are components of tensors of rank 2. Since $C_\mu\in\cl_\circledS(p,q)\T_1$, then (by Theorem \ref{LemmaDif}) $\partial_\mu C_\nu\in\cl_\circledS(p,q)\T_2$. So
$$
\partial_\mu C_\nu-\partial_\nu C_\mu-[C_\mu,C_\nu] \in\cl_\circledS(p,q)\T_2
$$
and, hence, $c_{\mu\nu}=0$,\,$d_{\mu\nu}=0$. Equality (\ref{CmuCnu}) is proved. Gauge invariance of equality
(\ref{CmuCnu}) is proved by the formula
$$
\partial_\mu\acute C_\nu-\partial_\nu\acute  C_\mu-[\acute C_\mu,\acute C_\nu]=
S^{-1}(\partial_\mu C_\nu-\partial_\nu C_\mu-[C_\mu,C_\nu])S.\qquad \rule{5pt}{5pt}
$$

\section{General solution of the primitive field equation}

In this section we find a general solution (up to elements of the center of Clifford algebra) of the primitive field equation (\ref{nik:eq}).

\begin{theorem}{Theorem}
Suppose that $n$ is a natural number and $C_\mu\in\cl_\circledS(p,q)\T_1$. Then the following two systems of equations are equivalent:
\begin{equation}
\partial_\mu h_\rho-[C_\mu, h_\rho]=0 \quad\Leftrightarrow\quad C_\mu=\sum_{k=1}^{\acute n} \mu_k \pi[h]_k ((\partial_\mu h^\rho) h_\rho),\label{resh}
\end{equation}
where $\acute n=n$ for even $n$, $\acute n=n-1$ for odd $n$ and
$
\mu_k=\frac{1}{n-(-1)^k(n-2k)}=\frac{1}{n-\lambda_k}.
$
\end{theorem}

Remark. Using formulas (\ref{svpr2}), we can rewrite general solution (\ref{resh})
of the primitive field equation in the following form (we use contractions and do not use projection operators):
\begin{eqnarray}
C_\mu=\sum_{k=1}^{n} \mu_k \sum_{l=0}^n b_{kl} F^l((\partial_\mu h^\rho) h_\rho)=\sum_{l=0}^{n}r_l F^l((\partial_\mu h^\rho) h_\rho),\quad r_l=\sum_{k=1}^n \mu_k b_{kl}\label{vid1}
\end{eqnarray}
in the case of even $n$ and
\begin{eqnarray}
C_\mu=\sum_{k=1}^{n-1} \mu_k \sum_{l=0}^{\frac{n-1}{2}} g_{kl} F^l((\partial_\mu h^\rho) h_\rho)=\sum_{l=0}^{\frac{n-1}{2}} s_l F^l((\partial_\mu h^\rho) h_\rho),\quad s_l=\sum_{k=1}^{\frac{n-1}{2}}  \mu_k g_{kl}\label{vid2}
\end{eqnarray}
in the case of odd $n$.

On the page \pageref{page:resh} we write explicit formulas for solution of the primitive field equation in the cases of small dimensions $n=2, 3, 4$.

\proof. Consider the decomposition of solution $C_\mu$ of system of equations (\ref{nik:eq})
\begin{equation}
C_\mu=\sum^n_{k=0}\pi[h]_k(C_\mu),\label{C:dec}
\end{equation}
where $\pi[h]_k(C_\mu)\in\cl[h]_k(p,q)\T_1$. Multiply the left side of equation (\ref{nik:eq}) by $h^\rho$ and consider the corresponding contraction (summation over index $\rho$): $h^\rho\partial_\mu h_\rho - h^\rho C_\mu h_\rho + h^\rho h_\rho C_\mu = 0$. Using formula (\ref{C:dec}) and formulas
$$
h^\rho h_\rho=n e,\quad h^\rho C_\mu h_\rho=\sum_{k=0}^n
h^\rho\pi[h]_k(C_\mu)h_\rho=\sum_{k=0}^n (-1)^k(n-2k)\pi[h]_k(C_\mu),
$$
we obtain
\begin{equation}
\sum_{k=0}^n(n-(-1)^k(n-2k))\pi[h]_k(C_\mu)=-h^\rho\partial_\mu
h_\rho=(\partial_\mu h^\rho)h_\rho.\label{Chformula}
\end{equation}
It easy to see that $n-(-1)^k(n-2k)=0$ holds for $k=0$, $\forall n$ and for $k=n$, odd $n$. From (\ref{Chformula}) we obtain required formula (\ref{nik:eq}) for $C_\mu$.

Now we shall prove that this expression for $C_\mu$ satisfies the primitive field equation.

Consider the following contractions $M^{a, s}_{(-1)^t}(U)$:
\begin{eqnarray}
M^{a,s}_1(U, h_\nu)&=&h^{\mu_1}\cdots h^{\mu_s} h^{\rho_1} \cdots h^{\rho_a} U h_{\rho_a} \cdots h_{\rho_1}h_\nu h_{\mu_s} \cdots h_{\mu_1},\nonumber\\
M^{a,s}_{-1}(U, h_\nu)&=&h^{\mu_1}\cdots h^{\mu_s}h_\nu h^{\rho_1} \cdots h^{\rho_a} U h_{\rho_a} \cdots h_{\rho_1} h_{\mu_s} \cdots h_{\mu_1}.\nonumber
\end{eqnarray}

We contract an arbitrary element $U\in\cl(p,q)$ over $a+s$ indices. An element $h_\nu$ is on the right if $t=0$ and on the left if $t=1$. The number $s$ is a distance between $h_\nu$ and the boundary of expression, the number $a$ is a distance between $h_\nu$ and the center of expression.

\begin{lemma}{Lemma} We have $M^{a,s}_{(-1)^t}(U, h_\nu)=-M^{a-1,s+1}_{(-1)^t}(U, h_\nu)+2M^{a-1,s}_{(-1)^{t+1}}(U, h_\nu).$
\end{lemma}

\proof. In the case $t=0$ we permute neighboring elements $h_\nu$ and $h_{\rho_1}$ using $h_{\rho_1}h_\nu=-h_\nu h_{\rho_1}+2\eta_{\nu \rho_1} e$ and obtain 2 another contractions from the statement. In the case $t=1$ we use $h_\nu h^{\rho_1}=-h^{\rho_1}h_\nu+2\eta^{\rho^1}_\nu e$. \rule{5pt}{5pt}

\begin{lemma}{Lemma} We have $M^{a, s}_{(-1)^t}(U, h_\nu)=\sum_{i=0}^a (-1)^i 2^{a-i} C_{a}^{a-i} M^{0, i+s}_{(-1)^{a-i+t}}(U, h_\nu).$
\end{lemma}

\proof. We use the method of mathematical induction (over index $a$). For $a=0$ we have $M^{0, s}_{(-1)^t}(U, h_\nu)=M^{0, s}_{(-1)^t}(U, h_\nu)$.
Suppose that this formula is valid for some $a$. Let us prove the validity of this formula for $a+1$. We have
\begin{eqnarray}
&&M^{a+1, s}_{(-1)^t}=-M^{a, s+1}_{(-1)^t}+2M^{a, s}_{(-1)^{t+1}}\nonumber\\
&&=-\sum_{i=0}^a (-1)^i 2^{a-i} C_{a}^{a-i} M^{0, i+s+1}_{(-1)^{a-i+t}}+2\sum_{i=0}^a (-1)^i 2^{a-i} C_{a}^{a-i} M^{0, i+s}_{(-1)^{a-i+t+1}}\nonumber\\
&&=\sum_{j=1}^{a+1}(-1)^{j}2^{a-j+1}C_{a}^{a-j+1}M^{0, j+s}_{(-1)^{a-j+1+t}}+\sum_{i=0}^a (-1)^i 2^{a-i+1} C_{a}^{a-i} M^{0, i+s}_{(-1)^{a-i+t+1}}\nonumber\\
&&=\sum_{i=1}^{a} (-1)^i 2^{a+1-i} (C_{a}^{a-i+1}+C_{a}^{a-i})M^{0, i+s}_{(-1)^{a-i+t+1}}+(-1)^{a+1} M^{0, a+1-s}_{(-1)^t}+2^{a+1}M^{0,s}_{(-1)^{a+t+1}}\nonumber\\
&&=\sum_{i=0}^{a+1} (-1)^i 2^{a+1-i} C_{a+1}^{a+1-i} M^{0, i+s}_{(-1)^{a+1-i+t}},\nonumber
\end{eqnarray}
where we use $C_n^{k+1}+C_n^{k}=C_{n+1}^{k+1}$ and use notation $M^{a,s}_{(-1)^t}(U, h_\nu)=M^{a,s}_{(-1)^t}$. \rule{5pt}{5pt}

We continue the proof of the theorem in the case of even $n$. Let us substitute formulas (\ref{vid1}) for $C_\mu$ in the primitive field equation:
$$\partial_\mu h_\nu=\sum_{l=0}^n r_l F^l((\partial_\mu h^\rho) h_\rho) h_\nu- \sum_{l=0}^n r_l h_\nu F^l((\partial_\mu h^\rho) h_\rho).$$

Using Lemmas, we obtain
\begin{eqnarray}
\partial_\mu h_\nu&=&\sum_{l=0}^n r_l F^l((\partial_\mu h^\rho) h_\rho) h_\nu- \sum_{l=0}^n r_l h_\nu F^l((\partial_\mu h^\rho) h_\rho)\nonumber\\
&=&\sum_{l=0}^n r_l(M^{l, 0}_{1}((\partial_\mu h^\rho) h_\rho, h_\nu)-M^{l,0}_{-1}((\partial_\mu h^\rho) h_\rho, h_\nu))\nonumber\\
&=&\sum_{l=0}^n r_l \sum_{i=0}^l (-1)^i 2^{l-i} C_l^{l-i}(M^{0, i}_{(-1)^{l-i}}((\partial_\mu h^\rho) h_\rho, h_\nu)-M^{0,i}_{(-1)^{l-i+1}}((\partial_\mu h^\rho) h_\rho, h_\nu)).\nonumber
\end{eqnarray}
We have
\begin{eqnarray}
&&M^{0, i}_{(-1)^{l-i}}((\partial_\mu h^\rho) h_\rho, h_\nu)-M^{0,i}_{(-1)^{l-i+1}}((\partial_\mu h^\rho) h_\rho, h_\nu)\nonumber\\
&&=(-1)^{l-i}h^{b_1}\cdots h^{b_i} ((\partial_\mu h^\rho) h_\rho h_\nu-h_\nu(\partial_\mu h^\rho) h_\rho) h_{b_i} \cdots h_{b_1}\nonumber\\
&&=(-1)^{l-i} F^i((\partial_\mu h^\rho) h_\rho h_\nu-h_\nu(\partial_\mu h^\rho) h_\rho)\nonumber
\end{eqnarray}
and
\begin{eqnarray}
&&(\partial_\mu h^\rho) h_\rho h_\nu-h_\nu(\partial_\mu h^\rho) h_\rho=(\partial_\mu h^\rho)( -h_\nu h_\rho+2\eta_{\nu\rho}e) +h_\nu h_\rho(\partial_\mu h^\rho)\nonumber\\
&&=-(\partial_\mu h^\rho)h_\nu h_\rho+2\partial_\mu h_\nu +(-h_\rho h_\nu+2\eta_{\rho\nu}e)(\partial_\mu h^\rho)\nonumber\\
&&=4\partial_\mu h_\nu -((\partial_\mu h_\rho)h_\nu h^\rho+h_\rho h_\nu(\partial_\mu h^\rho))=4\partial_\mu h_\nu -(\partial_\mu(h_\rho h_\nu h^\rho)-h_\rho \partial_\mu (h_\nu) h^\rho)\nonumber\\
&&=4\partial_\mu h_\nu -((2-n)\partial_\mu h_\nu -h_\rho \partial_\mu (h_\nu) h^\rho)=(2+n)\partial_\mu h_\nu+h_\rho (\partial_\mu h_\nu) h^\rho.\nonumber
\end{eqnarray}

Then
\begin{eqnarray}
&&M^{0, i}_{(-1)^{l-i}}((\partial_\mu h^\rho) h_\rho, h_\nu)-M^{0,i}_{(-1)^{l-i+1}}((\partial_\mu h^\rho) h_\rho, h_\nu)\nonumber\\
&&=(-1)^{l-i} F^i((2+n)\partial_\mu h_\nu+h_\rho (\partial_\mu h_\nu) h^\rho)\nonumber\\
&&=(-1)^{l-i}\sum_{m=0}^n \lambda_m^i(2+n +\lambda_m)\pi[h]_m(\partial_\mu h_\nu),\nonumber
\end{eqnarray}
where $\lambda_m=(-1)^m(n-2m)$. So
$$\partial_\mu h_\nu=\sum_{l=0}^n r_l \sum_{i=0}^l (-1)^l 2^{l-i} C_l^{l-i}\sum_{m=0}^n \lambda_m^i(2+n +\lambda_m)\pi[h]_m(\partial_\mu h_\nu),$$
where $r_l=\sum_{k=1}^n\mu_k b_{kl}=\sum_{k=1}^n \frac{1}{n-\lambda_k} b_{kl}$ and
$B=||b_{kl}||$ is inverse of Vandermonde matrix.

We change index $j=l-i$ and change the order of summation:
\begin{eqnarray}
\partial_\mu h_\nu&=&\sum_{m=0}^n (2+n +\lambda_m)\pi[h]_m(\partial_\mu h_\nu)\sum_{k=1}^n \frac{1}{n-\lambda_k}\sum_{l=0}^n   b_{kl}  (-1)^l \sum_{j=0}^l 2^{j} C_l^{j}\lambda_m^{l-j}\nonumber\\
&=&\sum_{m=0}^n (2+n +\lambda_m)\pi[h]_m(\partial_\mu h_\nu)\sum_{k=1}^n \frac{1}{n-\lambda_k}\sum_{l=0}^n   b_{kl}  (-1)^l (2+\lambda_m)^l.\nonumber
\end{eqnarray}

Further we consider the sum over $m$ starting with  $m=1$ because $\pi[h]_0(\partial_\mu h_\nu)=0$.

We have $-2-\lambda_m=\lambda_{m+(-1)^{m+1}}$, $1\leq m\leq n$. Indeed, in the cases of even and odd $m$ we have respectively
\begin{eqnarray}
-2-\lambda_m=-2-(n-2m)=-2-n+2m=-(n-2(m-1))=\lambda_{m-1}=\lambda_{m+(-1)^{m+1}},\nonumber\\
-2-\lambda_m=-2+(n-2m)=-2+n-2m=n-2(m+1)=\lambda_{m+1}=\lambda_{m+(-1)^{m+1}}.\nonumber
\end{eqnarray}

Using $\sum_{l=0}^n b_{kl}(\lambda_a)^l=\delta_{k, a},$ we obtain
\begin{eqnarray}
\partial_\mu h_\nu&=&\sum_{m=1}^n (2+n +\lambda_m)\pi[h]_m(\partial_\mu h_\nu)\sum_{k=1}^n \frac{1}{n-\lambda_k}\sum_{l=0}^n   b_{kl}  (\lambda_{m+(-1)^{m+1}})^l\nonumber\\
&=&\sum_{m=1}^n (2+n +\lambda_m)\pi[h]_m(\partial_\mu h_\nu)\sum_{k=1}^n \frac{\delta_{k, m+(-1)^{m+1}}}{n-\lambda_k}\nonumber\\
 &=&\sum_{m=1}^n \frac{(2+n +\lambda_m)\pi[h]_m(\partial_\mu h_\nu)}{n-\lambda_{m+(-1)^{m+1}}}=\sum_{m=1}^n \pi[h]_m(\partial_\mu h_\nu).\nonumber
\end{eqnarray}
This completes the proof of theorem for the case of even $n$.

Let us prove theorem in the case of odd $n$. In this case we have $\lambda_k=\lambda_{n-k},$ hence $\mu_k=\mu_{n-k}.$

We have
\begin{eqnarray}C_\mu&=&\sum_{k=1}^{\frac{n-1}{2}} \mu_k \pi[h]_{k, n-k}((\partial_\mu h^\rho) h_\rho),\qquad \pi[h]_{k, n-k}(U)=\pi[h]_k(U)+\pi[h]_{n-k}(U),\nonumber\\
C_\mu&=&\sum_{k=1}^{\frac{n-1}{2}} \mu_k \sum_{l=0}^{\frac{n-1}{2}}g_{kl}F^l((\partial_\mu h^\rho) h_\rho)=\sum_{l=0}^{\frac{n-1}{2}}s_lF^l((\partial_\mu h^\rho) h_\rho),\qquad s_l=\sum_{k=1}^{\frac{n-1}{2}} \mu_k g_{kl}.\nonumber
\end{eqnarray}

Substitute this expression for $C_\mu$ in the primitive field equation and obtain
$$\partial_\mu h_\nu=\sum_{l=0}^{\frac{n-1}{2}}s_lF^l((\partial_\mu h^\rho) h_\rho) h_\nu- \sum_{l=0}^{\frac{n-1}{2}}s_l h_\nu F^l((\partial_\mu h^\rho) h_\rho).$$

Similarly to the case of even $n$ we get
$$\partial_\mu h_\nu=\sum_{l=0}^{\frac{n-1}{2}} s_l \sum_{i=0}^l (-1)^l 2^{l-i} C_l^{l-i}\sum_{m=0}^\frac{n-1}{2} \lambda_m^i(2+n +\lambda_m)\pi[h]_{m, n-m}(\partial_\mu h_\nu),$$
where $s_l=\sum_{k=1}^{\frac{n-1}{2}}\mu_k g_{kl}=\sum_{k=1}^{\frac{n-1}{2}} \frac{1}{n-\lambda_k} g_{kl}$ and
$G=||g_{kl}||$ is inverse of Vandermonde matrix.

Further we consider the sum over $m$ starting with  $m=1$ because $\pi[h]_{0, n}(\partial_\mu h_\nu)=0$. We change index $j=l-i$ and change the order of summation:

\begin{eqnarray}
&&\partial_\mu h_\nu=\sum_{m=1}^\frac{n-1}{2}(2+n +\lambda_m)\pi[h]_{m, n-m}(\partial_\mu h_\nu)\sum_{k=1}^{\frac{n-1}{2}}\frac{1}{n-\lambda_k} \sum_{l=0}^{\frac{n-1}{2}} (-1)^l g_{kl} \sum_{j=0}^l  2^{j} C_l^{j} \lambda_m^{l-j}\nonumber\\
&&=\sum_{m=1}^\frac{n-1}{2}(2+n +\lambda_m)\pi[h]_{m, n-m}(\partial_\mu h_\nu)\sum_{k=1}^{\frac{n-1}{2}}\frac{1}{n-\lambda_k} \sum_{l=0}^{\frac{n-1}{2}} (-1)^l g_{kl} (2+\lambda_m)^l.\nonumber
\end{eqnarray}

Using $-2-\lambda_m=\lambda_{m+(-1)^{m+1}}$, $1\leq m\leq n$ and $\sum_{l=0}^{\frac{n-1}{2}} g_{kl}(\lambda_a)^l=\delta_{k, a},$
we get
\begin{eqnarray}&&\partial_\mu h_\nu=\sum_{m=1}^\frac{n-1}{2}(2+n +\lambda_m)\pi[h]_{m, n-m}(\partial_\mu h_\nu)\sum_{k=1}^{\frac{n-1}{2}}\frac{\delta_{k, m+(-1)^{m+1}}}{n-\lambda_k}\nonumber\\
&&=\sum_{m=1}^\frac{n-1}{2}\frac{(2+n +\lambda_m)\pi[h]_{m, n-m}(\partial_\mu h_\nu)}{n-\lambda_{m+(-1)^{m+1}}}=\sum_{m=1}^\frac{n-1}{2}\pi[h]_{m, n-m}(\partial_\mu h_\nu).\nonumber
\end{eqnarray}

So, in the case of odd $n$ theorem is also proved. \rule{5pt}{5pt}

In the case of small dimensions $n=2, 3, 4$ expressions $C_\mu$ from the formula (\ref{resh}) have the following explicit form.

In the case $n=2$ \label{page:resh}
\begin{eqnarray}
C_\mu&=&\sum_{k=1}^{2} \mu_k \pi[h]_k ((\partial_\mu h^\rho) h_\rho)=\frac{1}{2}\pi[h]_1((\partial_\mu h^\rho) h_\rho)+\frac{1}{4}\pi[h]_2((\partial_\mu h^\rho) h_\rho)\nonumber\\
&=&\frac{1}{2}(\partial_\mu h^\rho) h_\rho-\frac{1}{16}h^\alpha(\partial_\mu h^\rho) h_\rho h_\alpha-\frac{3}{32}h^\beta h^\alpha (\partial_\mu h^\rho) h_\rho h_\alpha h_\beta.\nonumber
\end{eqnarray}

In the case $n=3$
\begin{eqnarray}
C_\mu&=&\sum_{k=1}^{2} \mu_k \pi[h]_k ((\partial_\mu h^\rho) h_\rho)=\frac{1}{4}\pi[h]_1((\partial_\mu h^\rho) h_\rho)+\frac{1}{4}\pi[h]_2((\partial_\mu h^\rho) h_\rho)\nonumber\\
&=&\frac{1}{4}\pi[h]_{12}((\partial_\mu h^\rho) h_\rho)=\frac{3}{16}(\partial_\mu h^\rho) h_\rho-\frac{1}{16}h^\alpha(\partial_\mu h^\rho) h_\rho h_\alpha.\nonumber
\end{eqnarray}

In the case $n=4$
\begin{eqnarray*}
C_\mu &=&\sum_{k=1}^{4} \mu_k \pi[h]_k ((\partial_\mu h^\rho)h_\rho)= \frac{1}{6}\pi[h]_1((\partial_\mu h^\rho)h_\rho)\\
&+&\frac{1}{4}\pi[h]_2((\partial_\mu h^\rho)h_\rho)+\frac{1}{2}\pi[h]_3((\partial_\mu h^\rho)h_\rho)+\frac{1}{8}\pi[h]_4((\partial_\mu h^\rho)h_\rho)\\
&=&\frac{1}{4}(\partial_\mu h^\rho) h_\rho +\frac{67}{576}h^{\alpha}(\partial_\mu h^\rho) h_\rho h_\alpha+\frac{73}{2304}h^{\beta} h^{\alpha} (\partial_\mu h^\rho) h_\rho h_{\alpha} h_{\beta}\\
&-& \frac{19}{2304}h^{\gamma}h^{\beta} h^{\alpha} (\partial_\mu h^\rho)h_\rho h_{\alpha} h_{\beta}h_{\gamma}-\frac{25}{9216}h^{\delta}h^{\gamma}h^{\beta} h^{\alpha} (\partial_\mu h^\rho) h_\rho h_{\alpha} h_{\beta}h_{\gamma}h_{\delta}.
\end{eqnarray*}

\medskip

\noindent{\bf Conclusion.}
We have invented a class of {\it primitive field equations} (\ref{nik:eq}), which depend on real Lie algebra of Lie group $G$. The equation (\ref{nik:eq}) has gauge symmetry w.r.t. the Lie group $G$. Also we discuss related mathematical structures -- Lie groups and Lie algebras in Clifford algebra, tensor fields with values in Clifford algebra, Clifford field vectors, an algebra of $h$-forms and so on.
Also we develop techniques needed to solve primitive fields equations -- theory of $h$-forms, a method of calculation of projection operators onto vector subspaces of $h$-forms of different ranks using inverse of Vandermonde matrices, etc.
We give general solutions of primitive field equations. In particular, we give explicit formulas for solutions in cases of small dimensions $n=2,3,4$.

Primitive field equations are model of Yang-Mills equations with Clifford field vector as a current in right hand part \cite{TrMIAN}. A necessity to solve primitive field equations arises, in particular, in model equations of field theory
\cite{mybook:eng} and in the theory of Dirac equation on curved  pseudo-Riemannian manifolds  \cite{Mitsk}.
Presented in this article results on primitive field equations show us a direction for further investigation -- to find new classes of solutions of Yang-Mills equations.


\begin{thebibliography}{99}

\bibitem{atiyah} Atiyah~M., {\em Vector Fields on Manifolds}, Arbeitsgemeinschaft f\"{u}r Forschung des Landes Nordrhein-Westfalen, Heft \textbf{200}, 1970.

\bibitem{Benn:Tucker} Benn~I.~M., Tucker~R.~W., {\em An introduction to Spinors and Geometry with Applications in Physics}, Bristol, 1987.

\bibitem{Cornw}J.~F.~Cornwell, {\em Group theory in physics}, Cambridge Univ. Press, 1997.

\bibitem{Doran} C.~J.~L.~Doran, D.~Hestenes, F.~Sommen and N.~van Acker, {\em Lie Groups as Spin Groups}, J.Math.Phys., \textbf{34}(8), 3642-3669 (1993).
\bibitem{kahler} K\"{a}hler~E., Randiconti di Mat. (Roma) ser. 5, 21, 1962, 425.

\bibitem{Lounesto} Lounesto~P., {\em Clifford Algebras and Spinors.}  Vol. 239 / L.M.S. Lecture Notes. Cambridge: Cambridge Univ. Press, 306 pp., 1997.

\bibitem{mybook:eng}N.~Marchuk, {\em Field theory equations}, Amazon, CreateSpace open publishing platform, ISBN 9781479328079, 290 p., 2012.

\bibitem{TrMIAN} N.~G.~Marchuk, {\em On a field equation generating a new class of particular solutions to the Yang-Mills equations}, \textit{Tr. Mat. Inst. Steklova}, \textbf{285}, 207-220, (2014) [\textit{Proceedings of the Steklov Institute of Mathematics}, 2014, Vol. \textbf{285}, pp. 197-210.]

\bibitem{mass:gen} N.~Marchuk, {\em Mass generation mechanism for spin-(1/2) fermions
in Dirac-Yang-Mills model equations with a symplectic gauge
symmetry}, \textit{Nuovo Cimento Soc. Ital. Fis. B}, \textbf{125:10}
(2010), 1249-1256.

\bibitem{MarSh2008} N.~G.~Marchuk, D.~S.~Shirokov, {\em Unitary spaces on Clifford algebras}, \textit{Adv. Appl. Clifford Algebr.}, \textbf{18:2} (2008), 237-254.

\bibitem{MarShir:book} N.~G.~Marchuk, D.~S.~Shirokov, {\em Vvedenie v teoriyu algebr Klifforda} (in Russian), Fazis, Moskva, 2012, 590 pp.
    %Н.~Г.~Марчук, Д.~С.~Широков, Введение в теорию алгебр Клиффорда, М:, Фазис, 590 стр., 2012.

\bibitem{Mitsk} N.~V.~Mitskevich, {\em Fizicheskie polya v obschey teorii otnositelnosty}, (in Russian), Nauka, Moskva, 1969, 325 pages.
    %Н.~В.~Мицкевич, Физические поля в общей теории относительности, М:, Наука, 325 стр., 1969.

\bibitem{Shirokov} D.~S.~Shirokov, {\em A classification of Lie algebras of pseudo-unitary groups in the techniques of Clifford algebras}, \textit{Advances in Applied Clifford Algebras}, Volume \textbf{20}, Number 2, pp. 411-425, (2010).

\bibitem{p-adic}D.~S.~Shirokov, {\em On some relations between spinor and orthogonal groups}, \textit{p-Adic Numbers, Ultrametric Analysis and Applications}, Vol.\textbf{3}, No.3, pp.212-218, (2011).

\bibitem{OTP} D.~S.~Shirokov, {\em Extension of Pauli's theorem to Clifford algebras}, \textit{Dokl. Math.}, \textbf{84}:2 (2011), 699-701.

\end{thebibliography}
\end{document}